\documentclass[aps,showpacs,preprintnumbers,amsmath,amssymb,superscriptaddress]{revtex4}
\usepackage{amssymb}
\usepackage{CJK}
\usepackage{amsmath,amssymb,graphicx,bm}

\begin{document}
\begin{CJK*}{GBK}{}

\title{The thermal entropy density of spacetime}

\author{Rongjia Yang}
\email{yangrongjia@tsinghua.org.cn}
 \affiliation{College of Physical Science and Technology,
Hebei University, Baoding 071002, China}
\affiliation{Department of Physics, Tsinghua University, Beijing 100084, China}


\begin{abstract}
We introduce the notion of thermal entropy density, and first obtain the thermal entropy density of any arbitrary spacetime without firstly assuming a temperature or a horizon. The results indicate that gravity possesses thermal effects or therm entropy density possesses effects of gravity. The results also imply that, besides gravity, the thermal entropy density can also be geometrized.
\end{abstract}

\makeatletter
\def\@pacs@name{PACS: 04.20.Cv, 05.70.Ce, 04.70.Dy}
\makeatother

\maketitle
\end{CJK*}

\section{Introduction}
In Newtonian theory of gravity, when a particle falls freely in a gravitational field, the gravity is also the inertial force. This fact leads to the principle that the inertial mass and the gravitational mass are equivalent. This principle can be considered as the first principle of equivalence. Based on this principle, Einstein suggested that the gravitational force and inertial force are equivalent. This equivalence between gravitational force and inertial force can be considered as the second principle of equivalence on which general relativity bases.

Besides these two relations: the relation between gravitational mass and inertial mass, and the relation between gravitational force and inertial force, it was also known that there is a profound connection between gravity and thermodynamics, as implied by the work of Cocke \cite{Cocke}, Bekenstein \cite{Bekenstein}, Hawking \cite{Hawking}, Davies \cite{Davies}, and Unruh \cite{Unruh}. After these studies, Wald had shown that the entropy $S$ can be taken to be the Noether charge associated with the diffeomorphism invariance of the theory \cite{Wald,Iyer}. Jacobson revealed that Einstein equations can be derived from the
first law of thermodynamics \cite{Jacobson}. This attempt had been generalized to the modified gravity \cite{Eling,Elizalde,Brustein} and had been revisited in \cite{Makela} which based on a consideration of the properties of a very small, spacelike two-plane in a uniformly accelerating motion. It had been shown that the field equations in both general relativity and Lovelock theories can be expressed as a thermodynamic identity near the horizon in a wide class of spacetime \cite{Padmanabhan,Paranjape,Kothawala} (see a review \cite{Padmanabhan2010}). Recently, the connection between gravity and thermodynamics had been proved to be held in the dynamical spacetime \cite{Wu}. By using the maximum entropy principle to a charged perfect fluid, the generalized Tolman-Oppenheimer-Volkoff is derived, which provides a strong evidence for the fundamental relationship between general relativity and ordinary thermodynamics \cite{Gao}. In cosmological context, the Friedmann equation can be rewritten in the form of the first low of thermodynamics \cite{Danielsson,Frolov,Calcagni,Cai,Cai2007,Akbar,Gong,Sheykhi,Ge,Wu2008}. In \cite{Bamba}, it had been explicitly shown that the equations of motion for modified gravity theories of $F(R)$-gravity, the scalar-Gauss-Bonnet gravity, $F(\mathcal{G})$-gravity and the non-local gravity are equivalent to the Clausius relation in thermodynamics. In \cite{Verlinde}, Verlinde argued that gravity can be explained as entropic force. Recently, it was shown that the Einstein-Hilbert action can be constructed by minimizing free energy \cite{Bracken}.

All these investigations were carried out in special contexts, or based on some assumptions, such as Unruh temperature, the existence of horizon, null surfaces, the apparent horizon, and so on. The key point in Jacobson's analysis, for example, bases on three assumptions: the first low of thermodynamics holds, the temperature experienced by the observer is the Unruh temperature, and the heat flow through the past Rindler horizon was defined to be the boost-energy current carried by matter. Does the connection between gravity and thermodynamics holds in any arbitrary spacetime? Can the analysis be carried out without assuming a specific expression of temperature or horizon? The difficulties are that we can not find a general expression of temperature, or, can not define a horizon in any arbitrary spacetime. In this paper, we try to investigate the relation between gravity and thermodynamics without firstly assuming a temperature or a horizon. The result we obtain implies that gravity possesses thermal effects, or, thermal entropy density possesses effects of gravity.

\section{Thermal entropy density of spacetime}
In general relativity or thermodynamics, both the energy density $\rho$ and the pressure $p$ play important roles. In general relativity, the energy density and the pressure are contained in the stress-energy tensor. In thermodynamics, $\rho$ and $p$ are contained in the first law of thermodynamics. We can reasonably conjecture that there may exist a relation between gravitation and thermodynamics. So let's begin with the first law of thermodynamics in curved spacetime
\begin{eqnarray}
\label{1st}
dE=TdS-p dV.
\end{eqnarray}
where $E$ is the total energy and $S$ is entropy within the volume $V$, $T$ is the temperature and $p$ is the pressure of the perfect fluid, and $dV=\sqrt{h}d^3x$ with $\sqrt{h}$ the determinant of the spatial metric. Throughout this paper, we take $c=G=1$ and use metric signature $(-,+,+,+)$. For a very small volume, the energy density can be considered as unchanged, so Eq. (\ref{1st}) can be rewritten as
\begin{eqnarray}
\label{11st}
(\rho+p)dV=TdS\equiv T s dV,
\end{eqnarray}
where $s$ is the entropy density. To avoid the difficulty of finding a specific expression of temperature, we introduce the thermal entropy density defined as $\sigma\equiv T s$, and reexpress Eq. (\ref{11st}) as
\begin{eqnarray}
\label{ted}
\sigma=\rho+p.
\end{eqnarray}
For radiation, $\rho=\alpha T^4$ with $\alpha$ a constant, $p=\rho/3$, and $s=4\alpha T^3/3$. It is obvious that we have $\sigma\equiv T s=\rho+p$. If $p=w\rho$ with $w$ a constant, the thermal entropy density is proportional to the energy density, and the change of the thermal entropy density with time is also proportional to that of the energy density, $\dot{\sigma} \propto \dot{\rho}$, for $w>0$. For dust (or dark matter) with $p\simeq 0$, the thermal entropy density is just the energy density. For $w<0$, the thermal entropy increases when the energy density decreases, and vice versa. The energy density $\rho$ and the pressure $p$ can be observed, so dose the thermal entropy density. So it can be concluded that the thermal entropy density is a notion which is more comprehensive than the notion of the energy density in thermodynamics.

Secondly, let's look for the expression of $\rho+p$ in general relativity so as to find the relation between gravity and thermodynamics. Thank of Einstein's equations \cite{Wald1984}
\begin{eqnarray}
\label{E}
R_{\mu\nu}-\frac{1}{2}g_{\mu\nu}R+\Lambda g_{\mu\nu}=8\pi T_{\mu\nu},
\end{eqnarray}
and the stress energy tensor of the perfect fluid
\begin{eqnarray}
\label{T}
T_{\mu\nu}=g_{\mu\nu}p+(\rho+p)u_{\mu}u_{\nu},
\end{eqnarray}
we obtain
\begin{eqnarray}
\label{E1}
R-4\Lambda =- 8\pi (3p-\rho).
\end{eqnarray}
To obtain the expression of $\rho+p$ in general relativity, we must find another equation about $\rho$ or $p$. We use the $3+1$ Einstein equations to attain this goal.
Let $n^{\mu}$ be the unit normal vector field to the 3 dimension hypersurfaces $\Sigma$, then we have \cite{Wald1984,Gourgoulhon}

\begin{eqnarray}
n^{\mu}n^{\nu}R_{\mu\nu}+\frac{1}{2}R-\Lambda =8\pi \mathcal{E},
\end{eqnarray}
where $\mathcal{E}=\Gamma^2(\rho+p)-p$ with $\Gamma$ the Lorentz factor. According to the scalar Gauss relation, we obtain
\begin{eqnarray}
\label{3E}
\mathcal{R}+K^2-K_{ij}K^{ij}-2\Lambda =16\pi \mathcal{E},
\end{eqnarray}
where $\mathcal{R}$ is Ricci scalar of the 3 dimension hypersurfaces $\Sigma$, $K_{ij}$ is the extrinsic curvature tensor of $\Sigma$, and $K$ the trace of the $K_{ij}$.
Combining Eqs. (\ref{E1}) and (\ref{3E}), we obtain the expression of $\rho+p$ in general relativity
\begin{eqnarray}
\label{ted1}
\rho+p=\frac{1}{4\pi(4\Gamma^2-1)} \left[\mathcal{R}+K^2-K_{ij}K^{ij}-\frac{1}{2}R \right].
\end{eqnarray}
From Eq. (\ref{ted}) and (\ref{ted1}), one can easily get
\begin{eqnarray}
\label{ted2}
\sigma=\frac{1}{4\pi(4\Gamma^2-1)} \left[\mathcal{R}+K^2-K_{ij}K^{ij}-\frac{1}{2}R \right].
\end{eqnarray}
The 4 dimension Ricci scalar, $R$, can be decomposed as \cite{Gourgoulhon}
\begin{eqnarray}
R=\mathcal{R}+K^2+K_{ij}K^{ij}-\frac{2}{N}\mathcal{L}_m K-\frac{2}{N}D_iD^{i}N,
\end{eqnarray}
where $\mathcal{L}_m$ is the Lie derivative along $\mathbf{m}$ of any vector tangent to $\Sigma$, and $D_i$ is the Levi-Civita connection associated with the metric of the 3 dimension hypersurfaces $\Sigma$. Then we can express the thermal entropy density with 3 dimension spacial geometrical quantities as
\begin{eqnarray}
\label{ted3}
 \sigma &=& \frac{1}{8\pi(4\Gamma^2-1)} \times \\\nonumber
&\times & \left[ \mathcal{R}+K^2-3K_{ij}K^{ij}+\frac{2}{N}\mathcal{L}_m K+\frac{2}{N}D_iD^{i}N \right],
\end{eqnarray}
Eq. (\ref{ted2}) or (\ref{ted3}) is the most important result we obtained in this work. The left-hand side of the equation is a quantity concerned with thermodynamics, while the right-hand side of the equation is related to the geometrical quantities of the spacetime. Eq. (\ref{ted2}) or (\ref{ted3}) tells us that they are equivalent.
Recall the case in Newtonian theory of gravity, when a particle free falls in a gravitational field, the gravity is also the inertial force, one can see that the inertial mass and the gravitational mass are equivalent. Now, the energy density $\rho$ and the pressure $p$ are not only the source of therm but also the source of gravity. This fact leads to Eq. (\ref{ted2}) or (\ref{ted3}) which implies that gravity possesses thermal effects, or, thermal entropy density possesses effects of gravity. We note that Eq. (\ref{ted2}) or (\ref{ted3}) holds for perfect fluid only, the case for non-perfect fluid will be discussed elsewhere.

In co-moving coordinate, Eq. (\ref{ted2}) takes the form \cite{Yang}
\begin{eqnarray}
\label{teds}
\sigma=\frac{1}{24\pi}(4R_{\rm s}-3R).
\end{eqnarray}
where $R_{\rm s}=g_{11} R^{11}+g_{22} R^{22}+g_{33} R^{33}$. In FRW universe, the thermal entropy density of the spacetime is $-(\dot{H}-k/a^2)/4\pi$ with $H=\dot{a}/a$ the Hubble parameter and $k$ a constant. In radiation dominated era, according to Einstein equations, we have $-(\dot{H}-k/a^2)/4\pi=\rho+p=4\alpha T^{4}/3=Ts\equiv \sigma$ with $\alpha=8\pi^5k^4_{\rm B}/(15h^3)$ \cite{Weinberg}, so Eq. (\ref{ted3}) holds.

\section{Conclusion}
We have introduced the notion of thermal entropy density via the first low of thermodynamics, and related it with 3 dimension spacial geometrical quantities vis Einstein's equations. We have obtained the therm entropy density of any arbitrary spacetime without firstly assuming a temperature or a horizon, that is to say, gravity can possess thermal effects, or, therm entropy density can possess effects of gravity. The results also indicate that, besides gravity, the thermal entropy density can also be geometrized. This may shed light on the nature of gravity. The thermal entropy density of spacetime can be applied to discuss the gravitational collapse. Here we have discussed the case of perfect fluid only and leaved the case of non-perfect fluid for future investigations.

\begin{acknowledgments}
This study is supported in part by National Natural Science Foundation of China under Grant No. 11147028, Hebei Provincial Natural Science Foundation of China under Grant No. A2011201147, and Research Fund for Doctoral
Programs of Hebei University under Grant No. 2009-155.
\end{acknowledgments}

\bibliography{apssamp}

\begin{thebibliography}{99}

\bibitem{Cocke}
 W. J. Cocke, Ann. Inst. Henri Poincar\'{e} \textbf{2}, 283 (1965).

\bibitem{Bekenstein}
J. D. Bekenstein, Phys. Rev. D \textbf{7}, 2333 (1973).

\bibitem{Hawking}
 S. W. Hawking, Commun. Math. Phys. \textbf{43}, 199 (1975).

\bibitem{Davies}
 P. C. W. Davies, J. Phys. A \textbf{8}, 609 (1975).

\bibitem{Unruh}
 W. G. Unruh, Phys. Rev. D \textbf{14}, 870 (1976).

\bibitem{Wald}
 R. M. Wald, Phys. Rev. D \textbf{48}, R3427 (1993).

\bibitem{Iyer}
 V. Iyer and R. M. Wald, Phys. Rev. D \textbf{50}, 846 (1994).

\bibitem{Jacobson}
 T. Jacobson, Phys. Rev. Lett. \textbf{75}, 126 (1995).

\bibitem{Eling}
 C. Eling, R. Guedens, and T. Jacobson, Phys. Rev. Lett. \textbf{96}, 121301 (2006).

\bibitem{Elizalde}
 E. Elizalde and P. J. Silva, Phys. Rev. D \textbf{78}, 061501 (2008).

\bibitem{Brustein}
 R. Brustein and M. Hadad, Phys. Rev. Lett. \textbf{103}, 101301 (2009).

 \bibitem{Makela}
 J. Makela and A. Peltola, Int. J. Mod. Phys. D \textbf{18}, 669 (2009).

\bibitem{Padmanabhan}
 T. Padmanabhan, Class. Quan. Grav. \textbf{19}, 5387 (2002).

\bibitem{Paranjape}
 A. Paranjape, S. Sarkar, and T. Padmanabhan, Phys. Rev. D \textbf{74}, 104015 (2006)

\bibitem{Kothawala}
 D. Kothawala, S. Sarkar, and T. Padmanabhan, Phys. Lett. B \textbf{652}, 338 (2007)

\bibitem{Padmanabhan2010}
 T. Padmanabhan, Rep. Prog. Phys. \textbf{73}, 046901 (2010).

\bibitem{Wu}
 S. F. Wu, B. Wang, X. H. Ge, and G. H. Yang, arXiv:1109.0193.

 \bibitem{Gao}
 S. Gao, Phys. Rev. D \textbf{84}, 104023 (2011).

\bibitem{Danielsson}
 U. H. Danielsson, Phys. Rev. D \textbf{71}, 023516 (2005)


\bibitem{Frolov}
 A. V. Frolov and L. Kofman, J. Cosmol. Astropart. Phys. \textbf{05}, 009 (2003).

\bibitem{Calcagni}
 G. Calcagni, J. High Energy Phys. \textbf{0509}, 060 (2005).

\bibitem{Cai}
 R. G. Cai and S. P. Kim, J. High Energy Phys. \textbf{02}, 050 (2005).

\bibitem{Cai2007}
R. G. Cai, L. M. Cao, and N. Ohta, Phys. Rev. D \textbf{81}, 061501(R) (2010)

\bibitem{Akbar}
 M. Akbar and R. G. Cai, Phys. Rev. D \textbf{75}, 084003 (2007).

\bibitem{Gong}
 Y. Gong and A. Wang, Phys. Rev. Lett. \textbf{99}, 211301 (2007).

\bibitem{Sheykhi}
 A. Sheykhi, B. Wang and R. G. Cai, Phys. Rev. D \textbf{76}, 023515 (2007).

\bibitem{Ge}
 X. H. Ge, Phys. Lett. B \textbf{651}, 49 (2007).

\bibitem{Wu2008}
 S. F. Wu, B. Wang, and G. H. Yang, Nucl. Phys. B \textbf{799}, 330 (2008).

 \bibitem{Bamba}
 K. Bamba, C. Q. Geng, S. Nojiri, and S. D. Odintsov, Europhys. Lett. \textbf{89}, 50003 (2010).

\bibitem{Verlinde}
 E. P. Verlinde, J. High Energy Phys. \textbf{04}, 029 (2011).

\bibitem{Bracken}
 P. Bracken, arXiv:1111.5068

\bibitem{Wald1984}
 R. M. Wald, \emph{General Relativity}, The University of Chicago Press, 1984.

\bibitem{Gourgoulhon}
 E. Gourgoulhon, arXiv:gr-qc/0703035

\bibitem{Yang}
 R.-J. Yang, preparing.

\bibitem{Weinberg}
 S. Weinberg, Gravitation and Cosmology: Principles and Applications of the Greneral Theory of Relativity, John Wiley, 1972.

\end{thebibliography}

\end{document}